\documentclass[11pt]{article}

\usepackage{amssymb}
\usepackage{amsmath}
\usepackage{epsfig}

\usepackage{amsfonts}
\usepackage{graphicx}
\usepackage{color}
\newtheorem{theorem}{Theorem}[section]
\newtheorem{lemma}[theorem]{Lemma}

\newtheorem{definition}{Definition}[section]

\newcommand{\be}{\begin{equation} \label}
\newcommand{\ee}{\end{equation}}
\newcommand{\bea}{\begin{eqnarray}\label}
\newcommand{\eea}{\end{eqnarray}}
\newcommand{\bas}{\begin{eqnarray*}}
\newcommand{\eas}{\end{eqnarray*}}
\newcommand{\bit}{\begin{itemize}}
\newcommand{\eit}{\end{itemize}}

\newcommand{\TCP}{\mathsf{TCP}}
\newcommand{\NTCP}{\mathsf{NTCP}}
\newcommand{\Var}{\mathsf{Var}}
\newcommand{\BED}{\mathsf{BED}}
\newcommand{\tmax}{{t_{\small\mathsf{max}}}}
\newcommand{\Dmax}{{D_{\small\mathsf{max}}}}
\newcommand{\deff}{{d_{\small\mathsf{eff}}}}
\begin{document}
\title{A Stochastic Model for the \\ Normal Tissue Complication Probability\\ (NTCP) in Radiation Treatment of Cancer}
\author{
Theresa Stocks\footnote{University of Stockholm, theresa.stocks@gmx.de}
\and
Thomas Hillen\footnote{Centre for Math.\ Biology, 
University of Alberta,  thillen@ualberta.ca}
\and 
Jiafen Gong \footnote{University of Toronto, gjf199936097@gmail.com}
\and
Martin Burger\footnote{Institut f\"ur Numerische und Angewandte Mathematik and Cells in Motion Cluster of Excellence, Westf\"alische Wilhelms-Universit\"at (WWU) M\"unster, martin.burger@wwu.de}
}
\date{}
\maketitle
%
\begin{abstract}
The normal tissue complication probability (NTCP) is a measure for the estimated side effects of a given radiation treatment schedule. Here we use a stochastic logistic birth death process to define an organ specific and patient specific NTCP. We emphasise an asymptotic simplification which relates the NTCP to the solution of a logistic differential equation. This framework allows for a direct use of the NTCP model in clinical practice. We formulate, but do not solve, related optimization problems. \\

\end{abstract}

\section{Introduction}
The goal of radiotherapy is to deliver a sufficient radiation dose to the tumor to provide a high probability of cure while the surrounding healthy tissue is minimally damaged and left functionally and architecturally competent.  To achieve this goal it is necessary to have a method of estimating the probability of normal tissue complication. Quantitative measures for the expected negative side effects on healthy tissue are  called {\it Normal Tissue Complication Probabilities} (NTCP) (\cite{Lyman85, Niemierko93,Stavrev:2001:GMT}). In this paper we investigate appropriate mathematical formulations thereof.  

The mathematical formulation of a NTCP is similar to the formulation of the {\it tumor control probability} (TCP), which represents the probability that after a radiation treatment no cancer cell has survived in the irradiated domain. The aim of treatment is to achieve a TCP value that converges, or is close, to one. While the TCP is concerned with the damage to cancerous tissue, 
the damage of surrounding healthy tissue cells is not included in a TCP model. Hence here we develop a cousin model, the NTCP, and we use the existing TCP models as guidelines for the development of NTCP models for healthy tissue. 

The formulation of a useful NTCP model has many challenges. NTCP models must be patient and organ specific. Details of the organ's function, the microenvironment, the biochemical pathways, the geometric structure, and the radio sensitivities are needed. For each organ (and each patient), we need to estimate a critical size such that the organ can still function. 
It is impossible to achieve this in an one-fits-all approach. 
Here we strive to prepare the development by providing a mathematical framework in which organ specific details can be included in an NTCP model. At the same time we strive to find a model that is not over burdened with complicated mathematics, and rather allows for a simple inclusion in clinical practice. 

The radiation damage to healthy organs and possible organ failures are inertly stochastic events, which cannot be predicted with certainty. The language of NTCP must, therefore, come from stochastic processes.  In this paper we focus on the mathematical aspects and we show that 
 \begin{itemize}
 \item Logistic birth-death models can be used to define a treatment-, patient-, and organ-specific NTCP (see Section \ref{s:logistic}).
 \item The NTCP can be estimated through the solution of the mean field equations, which allows for an estimation of a maximal tolerable dose $\Dmax$ for each patient (see Section \ref{s:clinical}).
\item We formulate (but do not solve here) an organ and patient specific optimization problem for radiation treatment with side effects. We have to leave it to future research to identify the necessary parameters for each organ/patient, and to perform the optimization (see Section \ref{s:conclusions}).
 \end{itemize}
 
Since the TCP and NTCP are closely related, we first review some TCP models before we extend them to NTCP modelling. One important ingredient is an estimate for the survival fraction $S(d)$, given a radiation dose $d$. We will review the corresponding linear quadratic model (LQ-model) in Section \ref{s:TCP}, where we also  
review models for the TCP. In Section \ref{s:Lyman} and Section \ref{s:volume} we discuss the NTCP approaches of {\it Lyman}  and the {\it critical volume} 
 approach. Section \ref{s:logistic} is devoted to our derivation of a stochastic NTCP model based on a  stochastic logistic process \cite{Feller39,Allen03}. 
 It turns out that the mean field equations of this process play an important role. Firstly, the mean field equations are of the form of a standard logistic differential equation plus a perturbation which depends on the variance. If the variance is small, or if the carrying capacity is large, then the mean field is basically a logistic equation. Moreover, we show that the region where the NTCP becomes critical (i.e. NTCP$\approx 1$) can be approximated by a Heaviside function, where the location of the jump coincides with the location where the solution of the logistic differential equation falls below a critical level. This relation is surprising, since the NTCP is an intrinsic stochastic concept, but it can actually be estimated from a deterministic differential equation. The same relation arose in the computation of the TCP from Zaider and Minerbo. It was never spelled out in \cite{ZaiderMinerbo2000}, but it was shown in \cite{GongEt11} that the Zaider-Minerbo TCP can be computed from the solution of the mean field equations. In Section \ref{s:clinical} we motivate the use of a maximal tolerable dose, based on the organ at hand, the patient's radio-sensitivities and the treatment schedule used. 
We close with a Conclusion section {\ref{s:conclusions} where we formulate corresponding optimization and optimal control problems.

\section{Previous models of cell survival, TCP and NTCP}\label{s:TCP}

TCP values are often obtained from \textit{statistical models of cell-survival}. These are  models that result from long clinical trials and research. Their advantages are simplicity and effective data-fitting. This is one of the main reasons why they are of high practical relevance. Unfortunately, they over-simplify important processes, leaving out complex cell mechanisms like repopulation of the cells and their differing sensitivity to irradiation.\\ 
Another approach is TCP derived from \textit{cell-population models}. These are models that take the stochasticity in case of  a small number of existing cells into consideration. Here the change of the cell density is described by stochastic processes, for example birth-death processes \cite{ZaiderMinerbo2000, Hanin04,Hillen:2005:CCD,HaninEt01}.  In contrast to the previous method these models consider important cell dynamics very precisely. We will introduce the Zaider-Minerbo TCP in Section \ref{s:ZMTCP} and then extend it in Section \ref{s:logistic} to NTCP.

Several models have been proposed to derive an NTCP. We review the three most prominent approaches including the Lyman NTCP in Section \ref{s:Lyman}, the critical volume NTCP in Section \ref{s:volume}, and the use of the {\it Biological Effective Dose}  in Section \ref{s:BED}. First, however, we recall the modelling of the surviving fraction after radiation treatment.  

\subsection{The linear quadratic model and the hazard function}\label{s:LQ}
We use $d$ to denote a radiation dose per fraction in units of $Gy$, and we use $D$ to denote the total dose. $S(D)$ is the surviving fraction of the tumor cells and for one radiation dose $d$ we use the well established linear quadratic model (LQ-model) 
\[
S(d) = e^{-\alpha d - \beta d^2}.\]
The parameters $\alpha$ and $\beta$ are called the {\it radiosensitivities}, and they have been measured for most cancerous tissues in the literature. For fractionated treatment with $n$ fractions and $D=nd$, the LQ formula is applied recursively to give 
\begin{equation}\label{LQfrac}
S(D) = e^{-(\alpha + \beta d) D}.
\end{equation}
This model has been extended (see \cite{Fowler10}) to include tumor re-growth with doubling time $T_p$ and a re-growth delay $T_k$ as observed in many instances. The extended LQ-model reads 

\begin{equation}\label{extendedLQ}
S(D)=\exp\left(-\alpha  \Bigl[nd\Bigl(1+ \frac{d}{\alpha/\beta}\Bigr)-\frac{\ln 2\;(T-T_k)}{\alpha T_p}\Bigr]\right).
\end{equation}

The model manages to include the biological effects like repopulation and healing, however the complexity of our original model has increased greatly. To quote Fowler: ``{\it The LQ-model loses its innocence when a time factor is added }'' \cite{Fowler10}.
The formula consists  of five unknown parameters instead of one and this leads to massive problems from a medical point of view as the parameters are difficult to measure. \\

If more complex treatment schedules are considered, for example unequal dosage, combination of radioactive seeds and external beam radiation, accelerated treatments etc., we can use a differential equation approach to define the surviving fraction at time $t$ as 
\[ \frac{d S(t)}{dt} = -h(t) S(t),\]
with {\it hazard function} $h(t)$. The hazard function carries all the details of the treatment schedule and, as shown in \cite{GongEt11}, the choice of the hazard function is important. 
The general hazard function can be written as 
\[ h(t) = (\alpha + \beta \deff(t)) \dot D(t), \]
where $\dot D(t)$ denotes the dose rate and $D(t)$ the total dose as function of time. The term $\deff(t)$ depends on the tissue at hand, on the radiation schedule and on the underlying physical radiation damage model. It is worthwhile to compare and contrast different approaches that are used in the literature. The most common are (see also \cite{GongEt11}): 
\[\begin{array}{lll}
(a) & \deff = d & \mbox{fractionated treatments}\\
(b) & \deff = 2 D(t) &\mbox{Zaider-Minerbo \cite{ZaiderMinerbo2000}}\\
(c) & \deff = 2\int_{-\infty}^t e^{-\gamma(t-s)} \dot D(s) ds & \mbox{Leah-Catchside protraction factor}\\
(d) & \deff = 2 (D(t)-D(t-\omega)) & \mbox{finite interaction window of single strand breaks}
\end{array}
\]
The coefficient $\gamma>0$ describes the exponential repair rate of single strand breaks. The exponential decay term in (c) leads to a reduced interaction of single strand breaks that are timely far apart. The coefficient $\omega>0$ has a similar function as it describes a time window such that single strand breaks which occur in this time window can interact to produce a double strand break. In many cases $\omega \approx 6 h$. Details of the modelling of hazard functions can be found in \cite{GongEt11}. 
  
In our numerical examples later, we study constant radiation with dose rate $d$. 
Using the above notation we have $\dot D(t) = d$, $D(t) = dt$ and 
the above choices for $\deff$ can be computed as 
\begin{eqnarray}
(a) && \deff = d\nonumber\\
(b) && \deff = 2 d t \label{h:unif}\\
(c) && \deff = \frac{2 d}{\gamma}\nonumber\\
(d) && \deff = 2d\omega\label{h:frac} 
\end{eqnarray}
We see that choice (b) is increasing in time, while all other choices are constant. 
The reason is that in (b) it is intrinsically assumed that single strand breaks can always interact, no matter how timely far apart they have been generated. The authors believe that this choice leads to an over aggressive hazard function. However, this choice is in popular use in the literature, and we feel obliged to include it here. 
The expressions for the Leah-Catchside protraction factor in (c) and for the finite interaction window in (d) are equivalent with the choice $\gamma^{-1} = \omega$. 

To illustrate our NTCP method we will later consider two types of tissues with  two types of hazard functions:
\begin{itemize}
\item {\bf Tissue A:} Here we assume the tissue regenerates very quickly if damaged, but we also assume that single strand breaks persist very long. Hence we choose the Zaider-Minerbo form of $h(t)$ (\ref{h:unif}). 
\item {\bf Tissue B:} Here we consider slow repair, and assume that single strand breaks interact on a time scale of up to 6h, i.e. we choose the hazard function $h(t)$ from (\ref{h:frac}). 
\end{itemize}
The corresponding survival fractions for the cases (\ref{h:unif}) and (\ref{h:frac}) can be computed to be

\begin{eqnarray*}
S(t) = e^{-(\alpha + \beta dt)d t}&& \mbox{for } (\ref{h:unif})\\
S(t) = e^{-(\alpha + 2\beta d \omega)dt} && \mbox{for } (\ref{h:frac}).
\end{eqnarray*}

\subsection{TCP by Zaider and Minerbo}\label{s:ZMTCP}
To motivate the TCP model of Zaider and Minerbo \cite{ZaiderMinerbo2000} we consider a simple ordinary differential equation for the tumor population density $n(t)$: 
\begin{equation}
{\frac{dn(t)}{dt}}= (b-r(t))n(t), \quad n(0) =n_0,
\label{zellpopzm}
\end{equation} 
with $b$ as a constant birth rate and $r(t)$ the removal rate. The removal rate can be written as the sum of the natural death rate $d$, which is constant and the radiation dependent hazard function  $h(t)$ giving $r(t) = d+ h(t)$. 
The solution of this cell population model is given by
\begin{equation}
n(t)=n_0\exp\Bigl(bt-\int_{0}^{t}r(s)ds\Bigr),\label{losung}
\end{equation}
with $n_0$ indicating the initial number of tumor cells.\\
\newline
For a large initial  number of cancer cells, deterministic models are appropriate, because with the law of large numbers  stochastic events can be neglected and the number of cells converges to the mean number of cells. However, a successful therapy aims to diminish  the number of cancer cells and for low cell numbers the  deterministic formulation no longer applies. Hence we extend the model to include stochastic events via a \textit{birth death process}. 

\noindent Following Zaider and Minerbo \cite{ZaiderMinerbo2000} we let $P_i(t)$ be the probability that $i$ cells are alive at time $t$ with $i\in \mathbb{N}$. 
The corresponding Master equation for $P_i(t)$ that describes the change of cells is then given as:
\begin{equation}
{\frac{dP_{i}(t)}{dt}}= (i-1)bP_{i-1}(t)+(i+1)r(t) P_{i+1}(t)-i(b+r(t) )P_{i}(t) 
\label{dglzm}
\end{equation}
setting $P_{-1}(t)=0$ and with initial values $P_{n_0}(0)=1$ and $P_i(0)=0$ for $i\neq n_0$.
It can be easily checked that the  expected number of tumor cells $ n(t)=\sum_{i=0}^{\infty} iP_i(t) $ satisfies the above equation (\ref{zellpopzm}) given that the sum converges. Hence (\ref{zellpopzm}) appears as {\it mean field} model for the stochastic birth-death process (\ref{dglzm}).\\
 To obtain the TCP we calculate $P_0(t)$. This can be done by methods of generating function 
 (see \cite{ZaiderMinerbo2000}). Hence we obtain the {\it TCP formula of Zaider-Minerbo} as  \begin{equation}\label{ZMTCP}
\TCP_{ZM}(t)= P_0(t)= \left[1-{\frac{n(t)}{n_0+bn_0n(t)\int_{0}^{t}\frac{dr}{n(r)}}}\right]^{n_0}
\end{equation}
with birth rate $b\geq 0$ and removal rate $r(t)$. Here $n(t)$ is the solution (\ref{losung}) of the mean field equation (\ref{zellpopzm}).

This framework has been extended to include active and quiescent cell compartments by Dawson and Hillen in \cite{Hillen:2005:CCD}, non-Poissonian cell cycle times by Maler and Lutscher in \cite{MalerEt09}, and cancer stem cells by Gong \cite{Gongthesis}. These models follow the same basic principle of stochastic processes, but the resulting TCP formulas are much more complicated.  In this paper we base the NTCP- formulation on the Zaider- Minerbo approach, being aware that further generalizations to include cell cycle and stem cells might be needed in the future.

\subsection{The Lyman-Model for NTCP}\label{s:Lyman}
The first and simplest model for the NTCP is a model developed by Lyman in 1985 \cite{Lyman85}. According to  his paper ``{\it a good treatment plan delivers a high uniform dose to the cancerous volume and lower dose to the surrounding normal tissues}'', also uniformly distributed. To measure the harm of the radiation on a particular healthy tissue,  an organ specific tolerance dose $TD_i$ is used. These are the doses that would result in $i$ $\%$ complication probabilities after 5 years. The tolerance doses can be described by functions that depend on beam area or fraction of the organ treated. Knowing the tolerance dose of the whole organ the relation to the tolerance dose of a fractional part $v$ $\in$ [0,1] of this organ is given by
\[
TD_{i}(v)= TD_{i}(1)v^{-n}
\]
with $i\in [0,100]$ and $0<n\leq 1$ a fitted parameter \cite{Niemierko93}.\\
\newline
According to \cite{Lyman85} data implies that normal tissue complication probability is not only a function of the absorbed dose  but also depends on the percentage of the organ volume irradiated. Keeping the fractional volume $v$ fixed (hence we set $TD_{50}(v) = TD_{50}$), this results in a  sigmoidal-shaped$\footnote{real valued and differentiable function having either a non-negative or a non-positive first derivative which is bell shaped\\  https://www.princeton.edu/~achaney/tmve/wiki100k/docs/Sigmoid\_function.html}$ NTCP-curve dependent on the total dose $D$.
The formula introduced in \cite{Lyman85} is the integral of a normal distribution with mean value $\mu=TD_{50}$ and standard derivation approximated by $\sigma= mTD_{50}$ receiving
\[
\NTCP_{Lyman}(D)= \frac{1}{\sqrt{2 \pi}\sigma}\int_{-\infty}^{D} e^{-\frac{1}{2}(\frac{z-\mu}{\sigma})^2}dz.
\]
Here $m$ is a parameter that governs the slope of the function, obtained from fitting clinical data \cite{Niemierko93}. 
Rescaling by $t=\frac{D-TD_{50}}{m TD_{50}}$ the formula reduces to a standard normal distribution and we obtain 
\begin{equation}
\NTCP_{Lyman}(D)= \frac{1}{\sqrt{2 \pi}}\int_{-\infty}^{t} e^{-\frac{s^2}{2}}ds.\label{lyman}
\end{equation}
This NCTP-formula is completely determined by the three parameters, $TD_{50}(1), m$ and $n$. For fixed $n, m$ and for variable partial volume $v$ and total dose $D$, the NCTP becomes a surface as shown in Fig. \ref{lymann}. This simulation is run for irradiation with parameters for the heart as $TD_{50}(1)=41.9$, $n=0.5$ and $m=0.1$ \cite{Lyman85}. 
\newline 
\begin{figure}
	\centering
	\includegraphics[width=10cm]{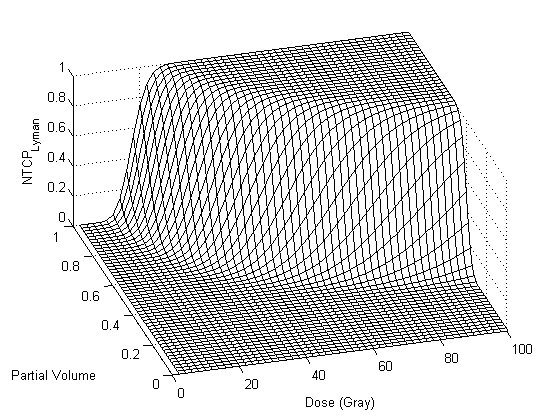}
	\caption{$\NTCP_{Lyman}$ for radiation of the heart as function of dose $D$ and fractional volume treated $v$.}\label{lymann}
	\end{figure}
	
 However, in daily practice the assumption of uniformly distributed dosage on tissue
can be relaxed with the emergence of  new technologies, such as CT scanned images,  making computerized treatment planning possible. These methods are able to generate images that allow for a 3D, non-uniformly distributed dose. 

\subsection{The Critical Volume NTCP}\label{s:volume}
Another deterministic model for NTCP is the so-called critical volume NTCP and was first introduced by Niemierko et al. in \cite{Niemierko93}. As detailed below, this model is closely related to the one of Lyman.   
A variety of organs can still function even when partially destroyed. This fact is called a \textit{parallel organ structure} and appears in organs such as the lung and kidney, because the undamaged parts work independently from the damaged ones \cite{Gongthesis}.\\
The critical volume model includes this tissue structure of the organ and is therefore more realistic. 
The smallest unit of an organ that is capable to perform biological functions is called a functional subunit (short: FSU) \cite{Stavrev:2001:GMT}. For example in the kidney the FSU are the renal tubes, in the liver it is the lobules, and the FSU of the lung are the acinuses. For the following we assume that an organ consists of $N$ FSUs being identical and uniformly distributed throughout  the organ. Each FSU consists of $N_0$ cells. To destroy one FSU all $N_0$ cells  have to be killed since we assume that a single cell is able to regenerate the FSU it is belonging to. The probability of damaging a FSU  after applying a dose $d$ is described by $P_{FSU}(d)$. To calculate this probability we use the complement of the LQ-model for describing the cell death within one FSU. Thus it is
\begin{equation}
P_{FSU}(d)=(1-e^{-(\alpha d+\beta d^2)})^{N_0}. \label{psfu}
\end{equation}
If we consider a fractionated treatment where the total dose $D$ is divided into $n$ fractions, not necessarily of equal dose, then (\ref{psfu}) becomes
\[
P_{FSU}(D)=(1-e^{-(\sum_{i=1}^{n}\alpha d_i+\beta d_i^2)})^{N_0}, 
\]
with $\sum_{i=1}^{n}d_i=D$.
Given these assumptions, the random variable  describing the probability that $i$ FSUs are killed after applying a dose $D$  is binomial distributed to the total number of  FSUs $N$ in the organ and the probability $P_{FSU}(D)$.
So we obtain
\[
P_{bin}(i)=\left( { N\atop i}\right)(P_{FSU}(D))^i(1-P_{FSU}(D))^{N-i}.
\]
For a large number $N$ of  FSUs, the distribution approaches the normal distribution by the central limit theorem, yielding
\[
P_{norm}(i)=\frac{1}{\sigma\sqrt{2\pi}}e^ {-\frac{1}{2}\left(\frac{i-\mu}{\sigma}\right)^2}
\]
with $\mu=\mathsf{E}(X)=NP_{FSU}$ and $\sigma^2=\Var(X)=NP_{FSU}(1-P_{FSU})$.
As we are looking for the probability that the normal tissue cannot function properly after radiation, we have to determine the probability that at least $R$ FSUs or more (up to all FSUs $N$) are damaged. Hence it is \[\NTCP_{cv}(D)=\sum\limits_{i=R}^N P_{bin}(i)\]
 and we obtain
\begin{equation}
\NTCP_{cv}(D)=\sum\limits_{i=R}^N \left( { N\atop i}\right) P_{FSU}(D)^i(1-P_{FSU}(D))^{N-i}.\label{njk}
\end{equation}
Moreover, we can approximate
\[
\NTCP_{cv}(D)=\sum\limits_{i=R}^N P_{bin}(i)\approx \int_{R}^{\infty} P_{norm}(i)di=\frac{1}{\sigma\sqrt{2\pi}} \int_{R}^{\infty}e^ {-\frac{1}{2}\left(\frac{i-\mu}{\sigma}\right)^2}di .
\]
We can easily see that after rescaling with the parameter $t= \frac{i-\mu}{\sigma}$, the critical volume NTCP has the same form as in Lyman, compare equation (\ref{lyman}).
\[
\NTCP_{cv}(D)= \frac{1}{\sqrt{2\pi}}\int_{\frac{R-\mu}{\sigma}}^{\infty}e^{\frac{-t^2}{2}}dt=\frac{1}{\sqrt{2\pi}}\int_{-\infty}^{\frac{\mu-R}{\sigma}}e^{\frac{-t^2}{2}}dt.
\]
\subsubsection*{Special cases}
Before we considered the general case that an organ is destroyed if $R$ or more FSUs are damaged. 
If an organ can only survive if all FSU are working properly then the organ has a so-called \textit{critical element} or \textit{serial architecture} and we get the special case   $R=1$. Equation (\ref{njk}) then reduces to the critical element NTCP
\[
\NTCP_{ce}(D)=\sum\limits_{i=1}^NP_{bin}(i)= 1-P_{bin}(0)=1-(1-P_{FSU})^N.
\]
Another special case is that the organ survives if at least one FSU survives ($R=N$). Hence equation (\ref{njk}) reduces to 
\[
\NTCP_{cv}(D)=P_{FSU}(D)^N.
\]
\subsubsection*{Inhomogeneous Dose Distribution}
So far we have assumed that the doses are homogeneously distributed on the tissue. We now want to relax this assumption and consider an heterogenic dose distribution on the irradiated tissue. Therefore we assume that the organ can be split into $k$ near-homogeneously irradiated  sub-volumes \cite{Niemierko93}, each one containing $K_i$ FSUs and receiving a dose $D_i$. The total number of killed FSUs, $N^{inhom}_{FSU}$, is then the sum of all killed FSU within the sub-volumes. This yields to 
\[
N^{inhom}_{FSU}= \sum_{i=1}^{k} K_iP^i_{FSU}(D_i)
\]
with $\sum_{i=1}^{R}K_i=N.$
The effective probability to kill one FSU with a heterogeneous dose distribution is given by
\[
P_{FSU}^{eff}=\frac{N^{inhom}_{FSU}}{N}.
\]

\subsection{NTCP as a function of BED}\label{s:BED}

Another concept for quality measure of radiation treatment is closely related to the improved LQ-model (\ref{extendedLQ}) and was firstly introduced by \cite{Fowler10}.
Instead of looking at the whole LQ-term we now only want to consider the exponent. We call the bracket term in (\ref{extendedLQ}) the {\it Biological Effective Dose} (short: BED) and receive the following definition:
\begin{equation*}
\BED_{\scriptscriptstyle{\alpha/\beta}}(n,d,T)=nd\left(1+ \frac{d}{[\alpha/\beta]}\right)-\frac{\ln 2\;(T-T_k)}{T_p\alpha}
\end{equation*}  
  with $n\in \mathbb{N}$ and $d,T_k,T_p,T \in \mathbb{R}^+$. It is a biologically effective dose for a tissue with a particular $\alpha/\beta$-ratio only. Fowler used the $\BED$ for treatment optimization in \cite{Fowler10}. He distinguished early and late side effects, whereby he assumed $\alpha/\beta=10$ for early responses and $\alpha/\beta=3$ for late responses. 

Assuming Poissonian statistics, the NTCP can be directly computed from the BED as 
\begin{equation*}
\NTCP_{BED}= \exp(-n_0\exp(-\alpha \BED))\sum_{k=0}^L \frac{(n_0\exp(-\alpha  \BED))^k}{k!}.
\end{equation*}

\subsection{Summary of previous models}
The NTCP models of Lyman and Niemierko are statistical models. Patient data on organ damage and survival are used to estimate parameters such as $TD_{50}, m, \sigma$. Later, in Section \ref{s:logistic} we will derive a mechanistic NTCP model that is based on the biological properties of the tissue at hand, allowing us to estimate the parameters based on organ tissue characteristics. \\

The biologically effective dose (BED) has been used for both, the tumor and the healthy tissue, and corresponding optimization problems have been studied (\cite{Fowler10}). 
The description of the TCP by Zaider and Minerbo provides a new level of detail as compared to the BED, for example. Any time-dependent treatment schedule can be included and the parameters are given from a birth-death process of tumor growth. However, so far, there was no cousin model for the NTCP which is based on an equally detailed description. The model which we develop next, will enable us to compare TCP and NTCP on equal grounds; and we will formulate a corresponding  optimization problem in the discussion Section \ref{s:conclusions}.  

\section{NTCP based on a stochastic logistic birth-death process}\label{s:logistic}
 
In this section we derive a NTCP model  from a stochastic logistic birth-death process. These are well known stochastic processes and detailed expositions can be found in the textbooks of Allen \cite{Allen03} and Nisbet and Gurney \cite{NisbetGurney}. The use of birth-death processes for NTCP is inspired by the  construction of a TCP from Zaider-Minerbo \cite{ZaiderMinerbo2000}. 
For our model we make the following assumptions:
\begin{enumerate}
\item Depending on the organ at hand, the entities of interest are either organ cells, or organ functional subunits. To keep the notation transparent, we will talk about cells in the following, but the model equally applies to functional subunits. We assume that all healthy tissue cells (or functional subunits)  in the irradiated domain are identical and independent throughout the organ. We plan in future work to extend this model and differentiate between stem cells and normal cells \cite{Stocks-stemcells}. 
\item Furthermore we assume that an organ works properly if more than $L$ cells (or functional subunits) exist.
\item For a small time increment $\Delta t$, the expression $\mu\Delta t$ denotes the probability of mitosis in a time interval $[t,t+\Delta t]$, where $\mu>0$ is the mitosis rate. We will assume  that the cell growth is limited by space and nutrition supply so the mitosis rate is dependent on an organ-specific carrying capacity $M$. An increasing number of cells therefore leads to a decreasing mitosis rate. Mathematically speaking we choose the mitosis rate as follows:
\begin{equation}
\mu_i=\left\{\begin{array}{rl} \mu(1-\frac{i}{M}),& \mbox{if } i=1,2,...,M \\ 0,& \mbox{otherwise} \end{array}\right.\label{birthrate}
\end{equation}
The carrying capacity $M$ refers to the organ size. If we count cell numbers then $M$ is usually a very large number ($\approx 10^9$) .
\item The term $r(t) = \rho + h(t)$ denotes the removal rate, where $\rho\geq 0$ denotes natural death of cells and the hazard function $h(t)$ death due to radiation (see (\ref{h:unif}) or (\ref{h:frac})). 
\end{enumerate}
We denote $P_i(t)$ as the probability that $ i\in \mathbb N$ normal cells are alive at time $t$. The probability that an organ cannot function properly anymore is then given by
\begin{definition}[NTCP  birth-death] The Normal Tissue Complication Probability based on a birth-death process is defined as
\begin{equation}\label{NTCP}
\NTCP_{bd}(t)=\sum\limits_{i=0}^{L} P_i(t).
\end{equation}
\end{definition}
The master equation for the probabilities $P_i(t)$ of the number of cells $X$ is given by  
\begin{equation}\label{master}
{\frac{dP_{i}(t)}{dt}}= (i-1)\mu_{i-1} P_{i-1}(t)+(i+1)r(t) P_{i+1}(t)-i(\mu_i+r(t) )P_{i}(t),
\end{equation}
with initial values $P_{n_0}(0)=1$ and $P_i(0)=0$ for $i\neq n_0$, \cite{Feller39,Allen03}.  
For the TCP we were only interested in the solution of $P_0(t)$. In contrast to that we are now interested in solving the system for $P_i(t)$ with $i= 0,...,L$. \\

The  mitosis rate as above guarantees that the number of normal tissue cells stays below or equal to the carrying capacity $M$.
\begin{lemma}\label{l:finite}{\rm (\cite{Feller39,Allen03}) }
Assume $\mu_i$ is given by (\ref{birthrate}). If $P_i(0)=0$ for $i\geq M+1$, then $P_i(t)=0$ for $i\geq M+1$, $\forall t>0$, i.e. the system (\ref{master}) is finite. 
\end{lemma}
Another interesting result shows that the mean field function $\mathsf{E}(X)$ (with $X$ denoting the random variable which describes the number of healthy tissue) obeys a logistic differential equation with a perturbation that depends on the variance. 
\begin{lemma}\label{t:meanfield}{\rm (\cite{Allen03} Formula (6.28), p. 246))} 
Assume $\mu_i$ is given by (\ref{birthrate}). Provided the series 
\begin{equation}
N(t)= \mathsf{E}(X)=\sum\limits_{i=0}^{\infty}iP_i(t) 
\end{equation}
converges, then $N(t)$ is the mean field function of system (\ref{master}) and satisfies a differential equation
\begin{equation}\label{meanfield}
\frac{dN(t)}{dt}=\mu N(t)\left(1-\frac{N(t)}{M}\right)-r(t)N(t)- \frac{\mu}{M} \Var(X),
\end{equation} 
where $\Var(X)$ is the variance of the normal tissue number and is defined as usual by $\Var(X)=\mathsf{E}((X-N(t))^2)$.
\end{lemma}
%
Let us provide some remarks on Lemma \ref{t:meanfield}: 
\begin{enumerate}
\item It is interesting to note that the perturbation term $\frac{\mu}{M} \Var(X)$ goes to zero for large carrying capacity $M$, or for small variance. In those cases we obtain the standard logistic differential equation for the expected number of cells $N(t)$. 
\item Since the variance is non negative, the mean field equation (\ref{meanfield}) is dominated by the logistic equation 
\begin{equation}\label{logistic}
 \frac{dZ}{dt} = \mu Z(t)\left(1-\frac{Z(t)}{M}\right)- r(t) Z(t) , 
 \end{equation}
i.e. $N(t)\leq Z(t)$ whenever they have the same initial condition $N(0)=Z(0)$ (a fact already known to Feller \cite{Feller39}, see also \cite{Allen03}).

\item We rescale the mean field equation (\ref{meanfield}) into a relative occupancy $y(t) := \frac{N(t)}{M}$. Then $y(t)$ satisfies
\begin{equation}\label{yeq}
\frac{dy}{dt} = \mu y (1-y) - r(t) y -\mu \Var Y, 
\end{equation}
where the random variable $Y$ is defined as $Y=X/M$. 
\end{enumerate}

These two previous results give us tools to compute the NTCP for the two complementary cases of $M$ is small and $M$ is large. If $M$ is small (say less than $1000$), then we benefit from Lemma \ref{l:finite}, the system of equations (\ref{master}) is of finite and manageable size, and we can use a direct numerical computation to solve it. This is done in the next Section \ref{s:smallM}.
On the other hand, if $M$ is large (larger than $1000$, say), then we can use an asymptotic method to approximate the NTCP as done in Section \ref{s:asymptotic}. We see that in this case the NTCP is basically given by the logistic differential equation (\ref{logistic}). If we compare these two methods (for $M=500$), we find that they coincide surprisingly well, suggesting that the logistic differential equation (\ref{logistic}) is appropriate  in computing the NTCP. We outline how it can be used in clinical practice in Section \ref{s:clinical}, where we also introduce the organ specific maximal tolerable dose $\Dmax$. 

\subsection{Numerical results for small $M$}\label{s:smallM}
After we have proved that the system of ODEs with $\mu_i$ is finite, we can now calculate the result numerically. We define $P(t)=(P_0(t),P_1(t),...,P_M(t))^T$  with $P_i(t)$ from (\ref{master}) and obtain a corresponding {\it forward Kolmogoroff equation} \cite{Allen03} 
\[
\frac{dP}{dt}=AP
\]
with the transition matrix A 
\[
A=  \begin{pmatrix}
 0 &  r(t)&0&...&0&0  \\
 0& -(\mu_1 +r(t))&2r(t)&...&&&\\
  \vdots &&&\ddots &&\\
 0&0&0&...&(M-1)\mu_{M-1}&-M(\mu_M+ r(t))
  \end{pmatrix}.
\]
For the initial values of the ODE system we chose a completely healthy organ at the beginning of treatment, i.e. $P_M(0)=1$ and $P_i(0)=0$ $\forall i\neq M$. Alternatively we can also consider partially damaged organs such that  $P_{n_0}(0)=1$ with $n_0<M$.  
We assume  continuous radiation per day with hazard function for uniform treatment (\ref{h:unif}) such that  \\
\[
\begin{split}
r(t)=& \rho +(\alpha +2 \beta dt)d\\
\mu_i =&\mu \Bigl(1-\frac{i}{M}\Bigr),\quad 0<i\leq M.
\end{split}
\]
For the simulation we used the valuesfrom Table \ref{t:para} as taken from \cite{Gongthesis}.
\begin{table}
 \begin{center}
 \begin{tabular}{|c|l|l|}\hline\hline
Parameter& Description& Value\\ \hline\hline
$\mu_A$ &birth rate fast in [$day^{-1}$]& 8.59 \\
$\mu_B$ &birth rate slow in [$day^{-1}$]& 0.07 \\
$\rho$ &natural death rate cells in[$day^{-1}$]& 0.03\\
M & carrying capacity & $500$\\
L& number of cells the organ needs to work properly& 0.05 M\\
$\alpha$ & sensitivity parameter in [$Gy^{-1}$] &0.06 \\
$\beta$&sensitivity parameter in [$Gy^{-2}$]&0.02 \\
$d$ & dose rate in [$Gy/day$]& 2.5, 3, 3.5 \\
$n_0$& initial number of cells& M \\
\hline\hline
\end{tabular}
\caption{Parameter values for a generic $\alpha/\beta=3$ healthy tissue, taken from \cite{Gongthesis}. \label{t:para}}
\end{center}
\end{table}
For the solution of the ODE-system we used the built-in MATLAB solver \textit{'ode45'}. Fig. \ref{plot1}  shows the simulated NTCP-curves for different constant dose rates (d= 2.5, 3.0, 3.5 Gy/day). On the left we show the NTCP curves for tissue type A with $\mu_A=8.59$ and hazard function (\ref{h:unif}), where on the right we have tissue type B with $\mu_B=0.07$ and hazard function (\ref{h:frac}). 
We observe that an increasing dose rate leads to shorter times before the healthy tissue is dysfunctional.\\

\begin{figure}[h!]
  \centerline{\includegraphics[width=7.2cm]{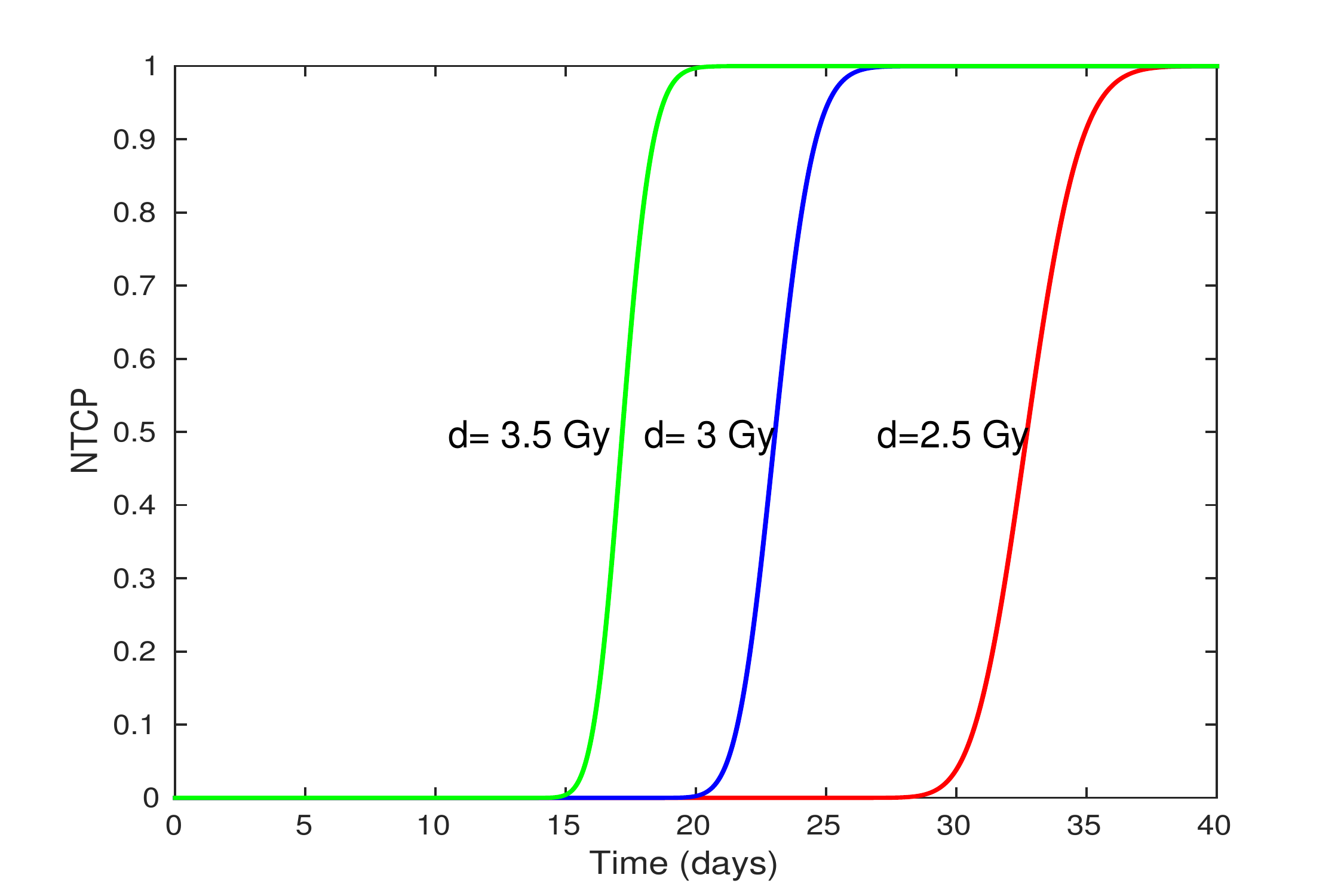}
  \includegraphics[width=7.2cm]{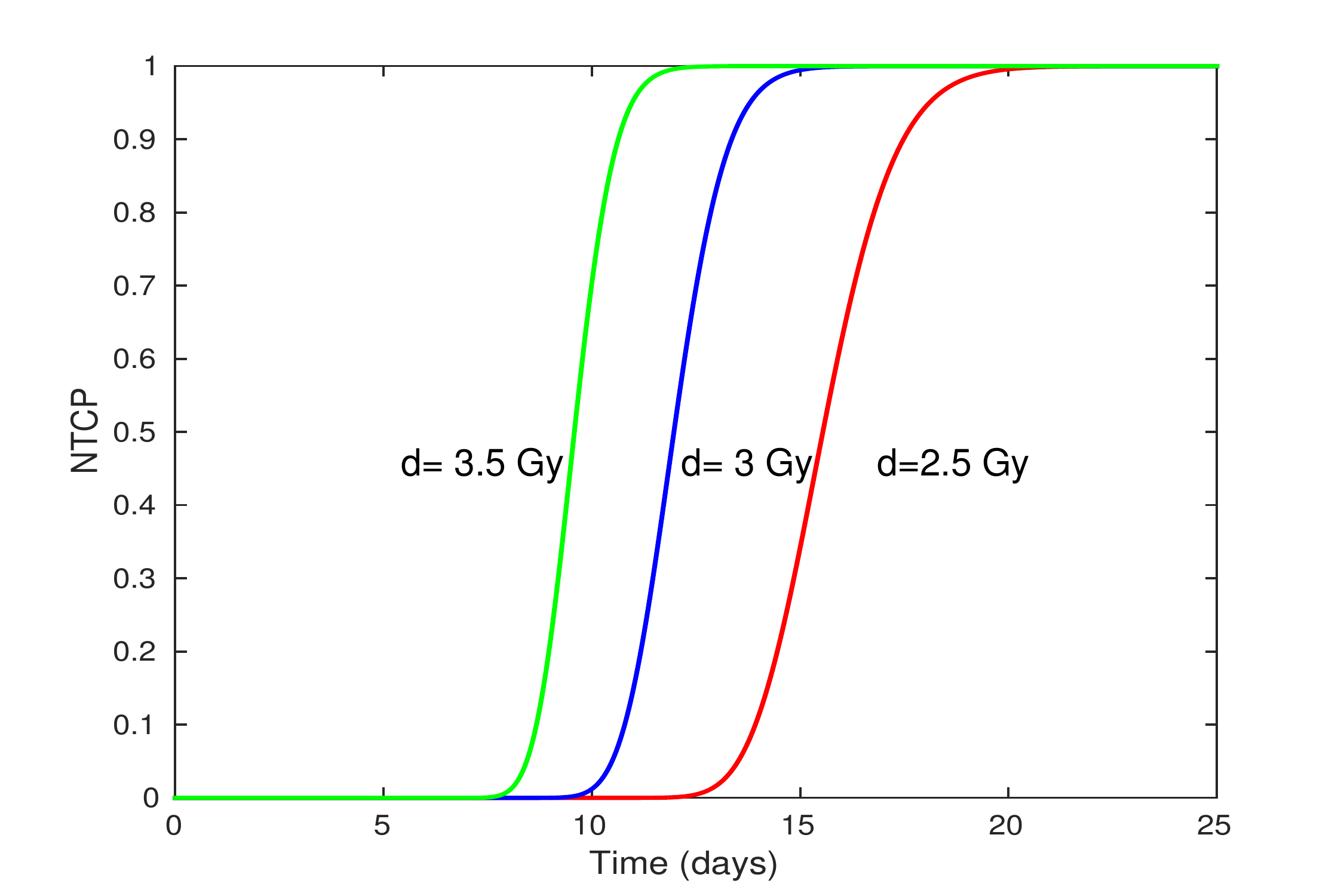}}
  \caption{NTCP curves as function of time (days) for constant radiation dose $d$ (Gy/day). Left: tissue tpe A, $\mu_A=8.59$ and hazard function (\ref{h:unif}); Right: tissue type B with $\mu_B=0.07$ and hazard function (\ref{h:frac})}
  \label{plot1}
\end{figure}


 \subsection{Asymptotics for large carrying capacity $M$}\label{s:asymptotic}
As mentioned before, in most cases, the carrying capacity $M$ will be large. Hence it is useful to consider the asymptotic limit of $M\to \infty$. In this case we still have the finite structure of the system of equations (\ref{master}), but the system is large, and asymptotic methods 
are a good alternative. 
We will use a rescaling argument to identify the location where the NTCP transfers from $0$ to $1$. It turns out that the mean field equation (\ref{logistic}) plays an important role for this transition.  \\

The question we find ourselves confronted with now is if there is an asymptotic so that the system of equations becomes independent of the size of the carrying capacity. Therefore the computational limits would not affect the simulations any longer and we could make capacity-independent predictions.
To achieve this aim we will re-parameterize the system of ODEs 
 (\ref{master}). As in the earlier numerical simulations  we assume that the initial number of cells is $n_0=M$. Therefore we get the initial values of the system $P_{M}(0)=1$ and $P_i(0)=0$ for $i\neq M$. For the parameterization we set
 \begin{equation}
P_i(t)= \frac{1}{M}\Phi \left(\frac{i}{M}, t\right)=  \frac{1}{M} \Phi (x,t)\label{defofphi}.
\end{equation}
where $x\in [0,1]$ is considered to be a continuous variable in the unit interval. The function $\Phi$ is a probability density, since $\Phi\geq 0$ and 
\begin{align*}
\int_0^1 \Phi(x,t)dx \approx \sum_{i=0}^M \frac{1}{M}\Phi(\frac{i}{M},t) =\sum_{i=0}^M P_i(t)=1.
\end{align*}
Hence we can modify our definition (\ref{NTCP}) of an NTCP as 
\begin{equation}
\NTCP_{\Phi}(t) =\int_0^l  \Phi(x,t)dx
\end{equation}
with $l=\frac{L}{M}\in[0,1]$. The expectation becomes
\[ N(t) = M\int_0^1 x \Phi(x,t) dx.\]
We now introduce the rescaling into the master equation (\ref{master}), where we have
\[
\frac{i}{M}=x,\quad \Delta x=\frac{1}{M},\quad \Delta x M=1,\quad i+1=(x+\Delta x)M, \quad i-1= (x-\Delta x) M\label{equations}
\]
and for the rescaled mitosis rate (\ref{birthrate}) we obtain
\begin{equation}
\tilde \mu(x)=\left\{\begin{array}{rl} \mu(1-x),& \mbox{if } x\in[0,1]\\ 0,& \mbox{otherwise}.\label{birthratescale} \end{array}\right.
\end{equation}
Using this and (\ref{defofphi}) for (\ref{master}) we obtain by Taylor-expansion with increment $\Delta x=\frac{1}{M}$  (see also \cite{NisbetGurney} eq. (6.2.18) on page 173) that 
{\allowdisplaybreaks 
\begin{align*}
\frac{\partial}{\partial t}\Phi(x,t)
=&-\frac{\partial}{\partial x}\bigl[(\mu x(1-x)+ r(t)x)\Phi(x,t)\bigr]\\
 & -\frac{\Delta x}{2}\frac{\partial^2}{\partial x^2}\bigl[(\mu x(1-x) + r(t) x) \Phi(x,t)\bigr]\\
 & + O(\Delta x^{-2})
\end{align*}}
If $\Delta x$ is small enough, i.e. $M$ large enough, then we can 
consider the leading order term of the above expansion. We obtain a hyperbolic partial differential equation
\begin{equation}
\frac{\partial}{\partial t}\Phi(x,t)=-\frac{\partial}{\partial x}\Bigl[(\mu x(1-x)- r(t)x)\Phi(x,t)\Bigr]\label{pdezaider}
\end{equation}
with initial values 
\begin{align*}
\Phi(1,0)&=\Phi \Bigl(\frac{M}{M}, 0\Bigr)=MP_M(0)=M. \\
\Phi(x,0)&=\Phi \Bigl(\frac{i}{M}, 0\Bigr) =MP_i(0) =0 \quad\text{for}\quad x\in[0,1).\\
\end{align*}
For $M\to\infty$ the above initial condition appears as a Dirac delta distribution 
\[\Phi(x,0) = \delta_1(x).\] 

We solve the PDE (\ref{pdezaider}) analytically using the method of characteristics. Expanding the spatial derivative we obtain 
\[
\frac{\partial}{\partial t}\Phi(x,t)+\Bigl[\mu x(1-x)-r(t)x\Bigr]\frac{\partial}{\partial x}\Phi(x,t) +(\mu(1-2x)-r(t)) \Phi(x,t) = 0.
\]
This hyperbolic PDE has the characteristic equations
 \begin{align} 
  \frac{dx}{dt} &= \mu x(1-x)-r(t)x& \quad x(0)&=x_0\label{chara1}\\
 \frac{d\Phi}{dt} &= -(\mu(1-2x)-r(t))\Phi, & 
\Phi(x,0)&=\delta_1(x).
 \label{chara2}
 \end{align}
With the initial value of $\Phi$ being a Dirac delta distribution and (\ref{chara2}) being linear in $\Phi$, we expect that  $\Phi(x,t)=\delta_{x(t)}$ is a weak solution of (\ref{pdezaider}). Here $x(t)$ is the solution of (\ref{chara1}) with the initial value $x_0=1$. See \cite{Evans98} for a definition of a weak solution. 
\begin{definition}
$\Phi\in \mathcal{D}(\Omega)=\mathbb{C}_0^{{\infty}^*}(\Omega)$ with $\Omega=[0,1]$ is a weak solution of (\ref{pdezaider}), if \begin{align*}
 \frac{d}{dt} \langle\Phi,\zeta \rangle=-\langle(\mu x(1-x)-r(t)x)\Phi(x,t) ,\frac{\partial}{\partial x}\zeta\rangle
\end{align*} for all $\zeta\in \mathbb{C}_0^\infty (\Omega)$.
\end{definition}
\vspace{0.2cm}
 In the following we will use the integral notation for the scalar product and write $\int \Phi(x,t)\zeta(x)dx$ instead of $\langle\Phi,\zeta\rangle$.
We obtain the following theorem:
\vspace{0.2cm}
\begin{theorem}\label{t:delta}
 Let $x(t)$ be the solution of (\ref{chara1}) with the initial value $x(0)=1$. Then  $\Phi(x,t) = \delta_{x(t)}$ is a weak solution of the PDE system (\ref{pdezaider}).\label{theorem1}
\end{theorem}
\textit{Proof.}
\begin{align*}
&\ &\frac{d}{dt}\int \Phi(x,t) \varphi(x)dx &= \int\frac{\partial}{\partial t}\Phi(x,t) \varphi(x) dx\\
&\  & &= -\int\left( \int\frac{\partial}{\partial t}\Phi(z,t)dz\right)\frac{\partial}{\partial x} \varphi(x) dx\\
& \ &  &= -\int \left(\int -\frac{\partial}{\partial z}((\mu z (1-z)-r(t)z)\Phi(z,t))dz\right)\frac{\partial}{\partial x} \varphi(x) dx\\
&\ & &=\int (\mu x (1-x)-r(t)x)\Phi(x,t)\frac{\partial}{\partial x}\varphi(x) dx\\
\intertext{where we used partial integration and (\ref{pdezaider}). Hence we obtain for $\Phi(x,t)=\delta_{x(t)}$}
&\ &\frac{d}{dt}\int \Phi(x,t) \varphi(x)dx &=\int (\mu x (1-x)-r(t)x)\Phi(x,t)\frac{\partial}{\partial x}\varphi(x) dx\\
&\Leftrightarrow & \frac{d}{dt}\varphi(x(t))&= (\mu x(t)(1-x(t))-r(t)x(t))\frac{\partial}{\partial x}\varphi(x(t))
\intertext{which is always true since}
&\ &\frac{d}{dt}\varphi(x(t))&= \frac{dx(t)}{dt} \frac{d\varphi(x)}{dx} =(\mu x(t) (1-x(t))-r(t)x(t))\frac{\partial}{\partial x}\varphi(x(t))
\end{align*}
with (\ref{chara1}). Hence $\delta_{x(t)}$ is a weak solution of (\ref{pdezaider}).\hfill $\square$\\
\newline

\subsection{Comparison of the two methods for small and large $M$}

We now want to compare  this asymptotic result with the numerical solution from the previous section, cp. Fig.\ref{plot1}. Using the solution of Theorem \ref{t:delta} we find 
\[
\NTCP_{\Phi}(t) =\int_0^l  \delta_{x(t)}(x)dx
\]
with $l=\frac{L}{M}\in[0,1]$. This integral is either $0$ if $x(t)>l$ or $1$  if $x(t)<l$. Hence the $\NTCP_\Phi$ function is a heavyside function that jumps at $t=x^{-1}(l)$ from 0 to 1. The following Figure \ref{f:char} shows the solutions of the characteristic ODE (\ref{chara1}), $x(t)$, (dotted)  for particular irradiation doses, the threshold value $l$ that indicates permanent damage on the healthy tissue (magenta) and the resulting $\NTCP_\Phi$- functions, which jump at the intersection point of cell  density function and tolerance value. The values we used for these simulations are the same values we used for Fig. \ref{plot1}, with tissue A on the left and tissue B on the right. 
\begin{figure}[h!]
  \centering
      \includegraphics[width=6.9cm]{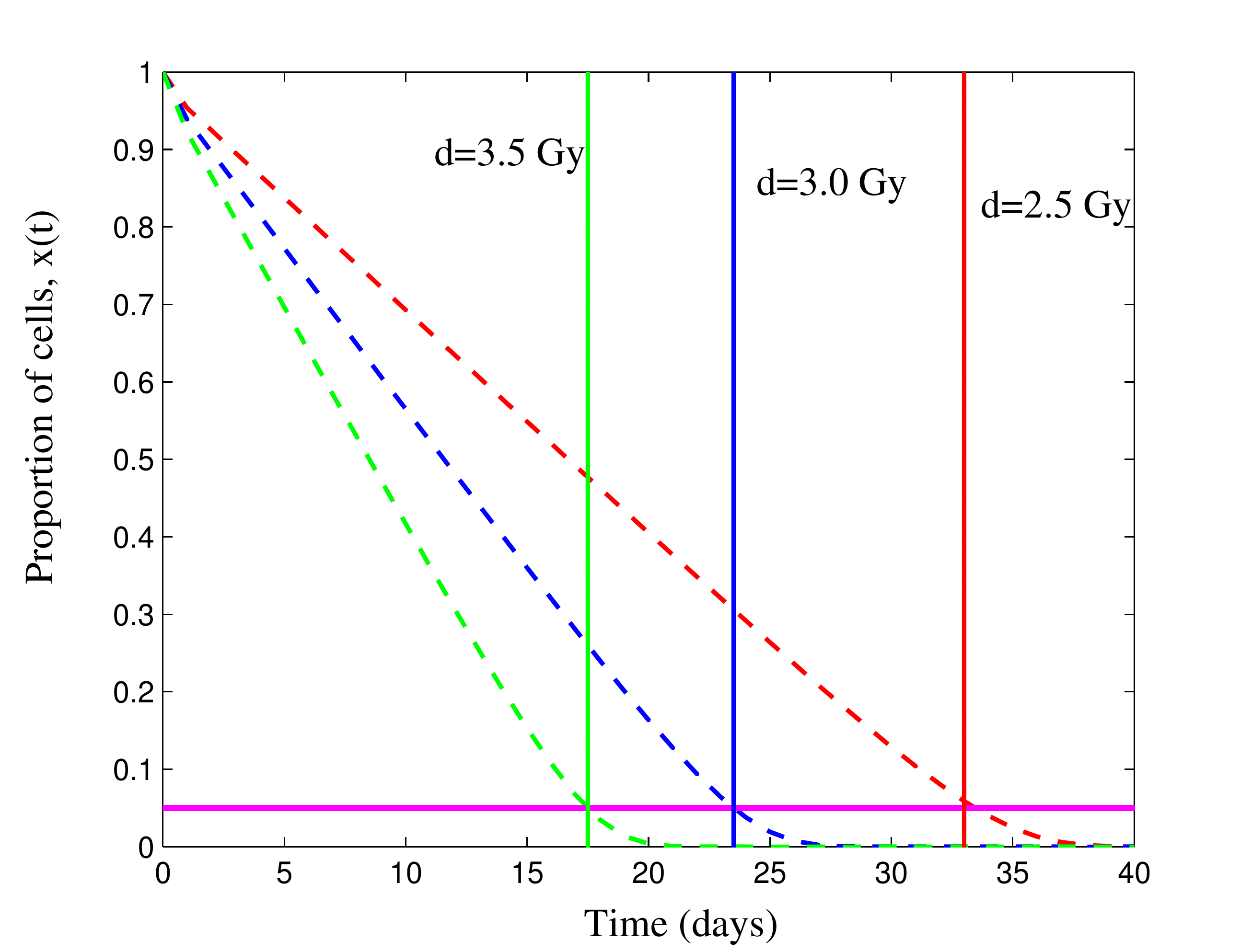}
      \includegraphics[width=7.7cm]{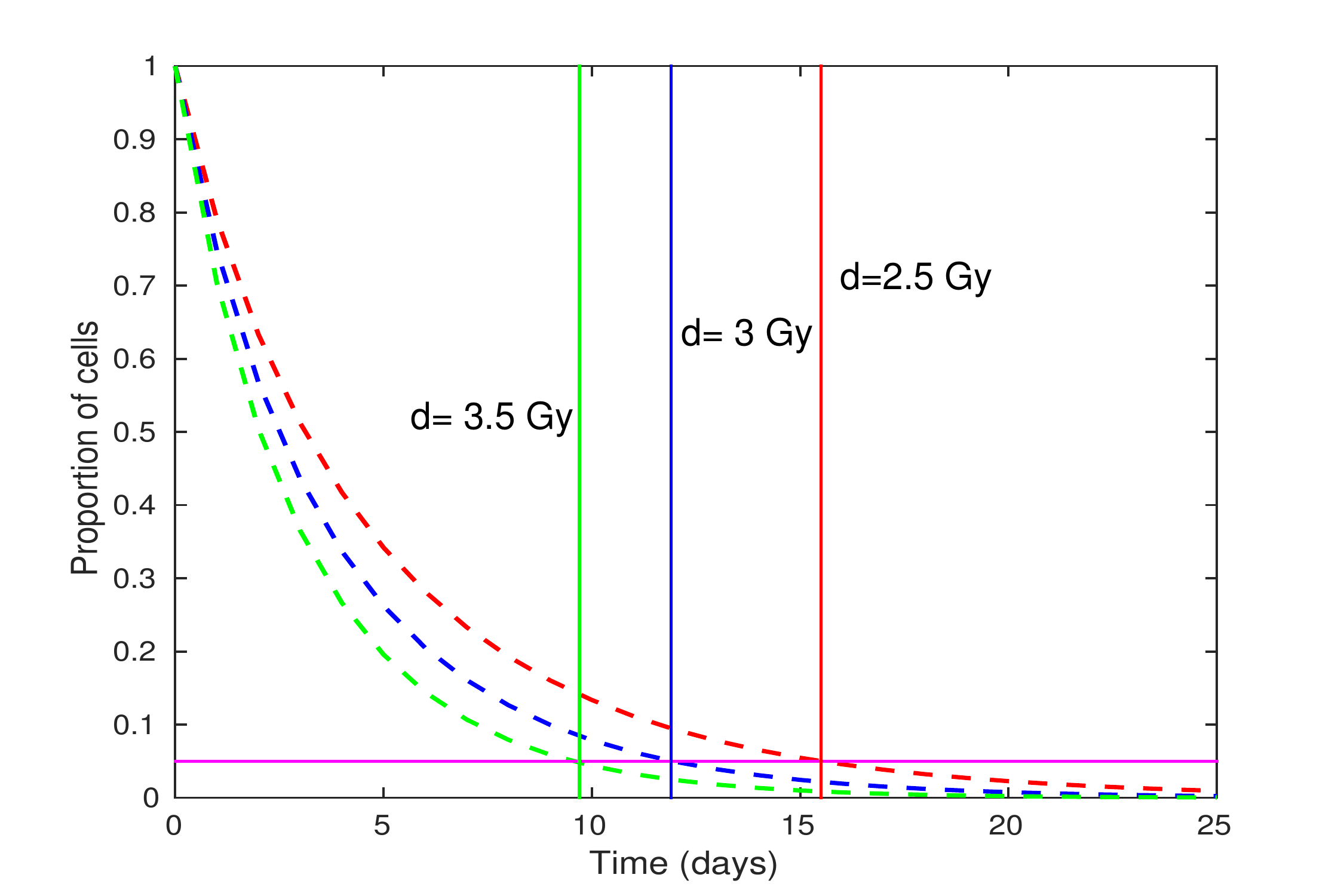}
  \caption{The dashed lines show the characteristics $x(t)$ for the three treatment dosages of 3.5, 3.0  and 2.5 Gy/day, for tissue A on the left and tissue B on the right.  They start at $x(0)=1$ and decrease until they intersect the vertical line of $x=L$. At that intersection the NTCP jumps from 0 to 1, indicated by a vertical line in the same color. \label{f:char}}
  
\end{figure}

In Figure 
\ref{compareNTCP} we compare the approximate NTCP curves with the ones from the numerical approach of Section \ref{s:smallM} and we see the transition regions coincide extremely well. We tried many more combinations of fast and slow regenerating tissues and various choices of hazard functions (not shown) and the correspondence of the two NTCP methods was always very good. Hence we succeeded in obtaining a carrying capacity independent formulation of the NTCP.   

\begin{figure}[h!]
  \centering
      \includegraphics[width=6.9cm]{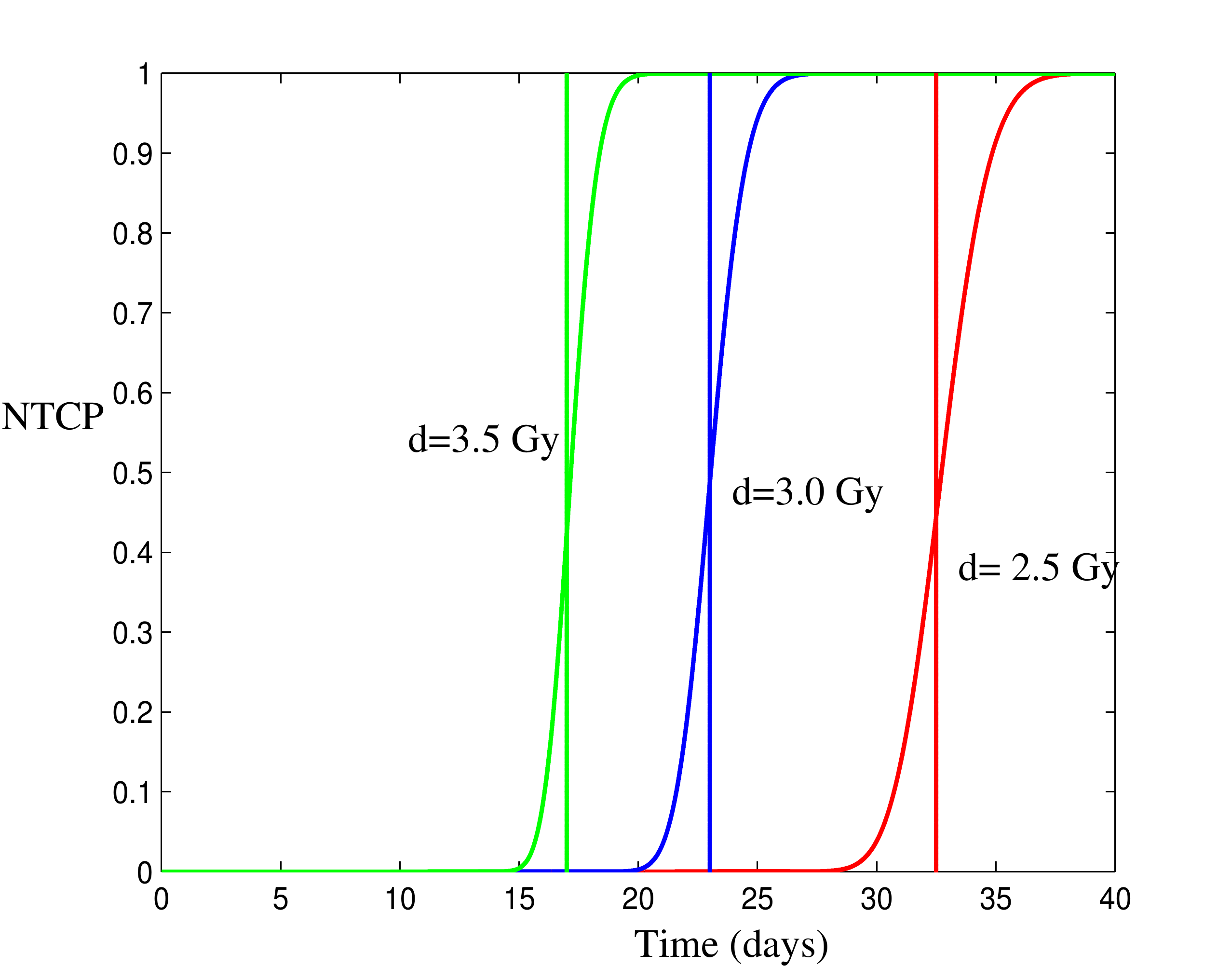}
      \includegraphics[width=7.7cm, height=5.5cm]{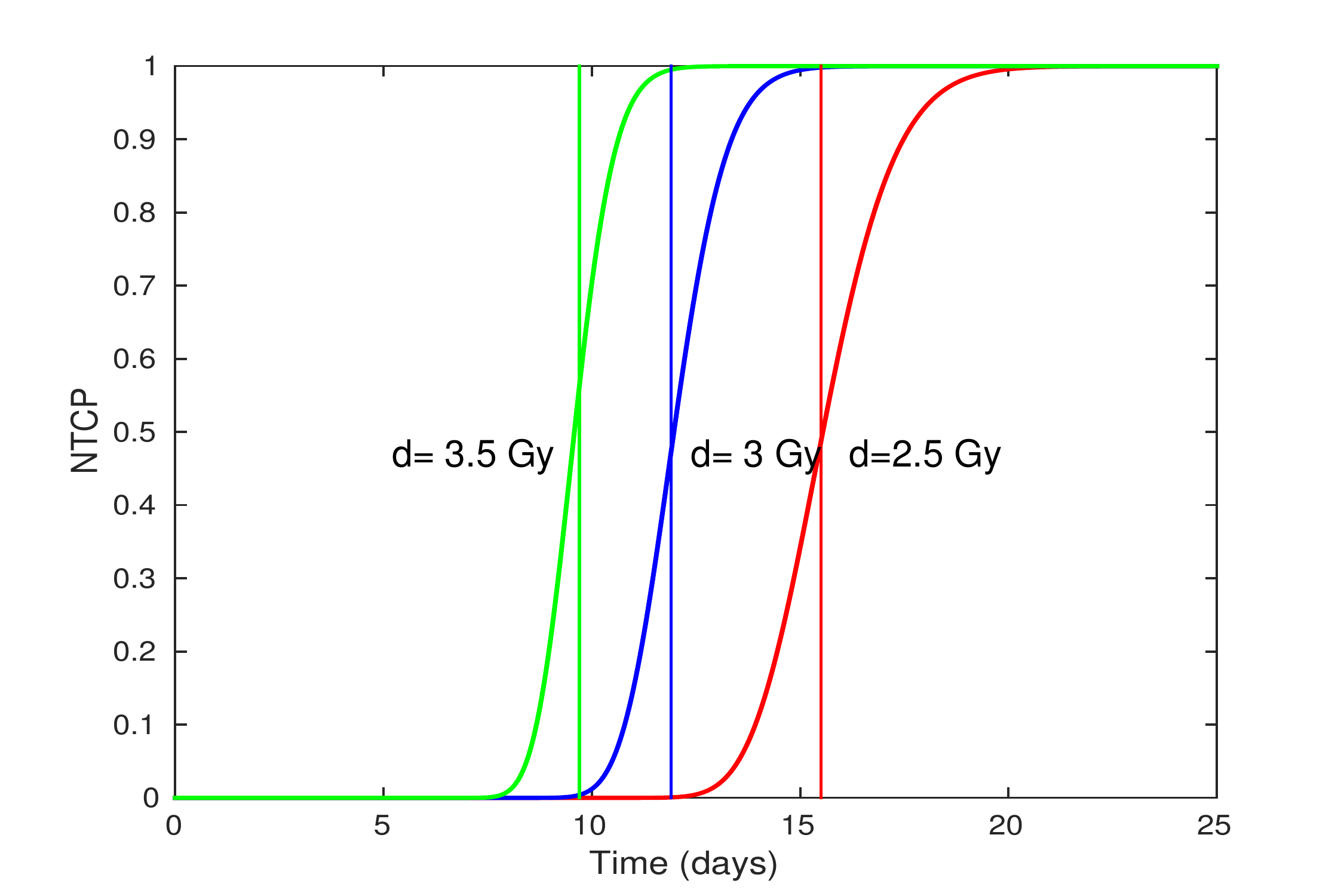}
  \caption{Overlay of the continuous NTCPs from Figure \ref{plot1} and the jumps from the asymptotic analysis, Figure \ref{f:char}. Left tissue type A, right tissue type B.\label{compareNTCP}}
  
\end{figure}
\section{Clinical significance and the maximal tolerable dose}\label{s:clinical}

Concerning the practical use of this NTCP, we propose an algorithm which is centered around the logistic differential equation (\ref{logistic}) which we rewrite for convenience:
\begin{equation}\label{logi2}
\frac{d Z}{dt}(t) = \mu Z(t)\left(1-\frac{Z(t)}{M}\right) - r(t) Z(t).
\end{equation}
We have seen in Lemma \ref{t:meanfield} that this equation approximates the mean field equations (\ref{meanfield}) for large carrying capacity $M$ or small variance. Moreover, if we consider the relative abundance $z(t) := Z(t)/M$, then we obtain
\[ \frac{d z}{dt}(t) = \mu z(t) (1-z(t)) - r(t) z(t),\]
which coincides with the characteristic equation (\ref{chara1}) that was used in the asymptotic method for large $M$. In Section \ref{s:asymptotic} we found that the $\NTCP_\Phi$ jumps from $0$ to $1$ exactly when the characteristic $x(t)$ meets the threshold value $l$. Transforming back to the original quantities $Z(t)$, an equivalent condition is 
\begin{equation}
Z(\tmax) = L,
\end{equation}
where $L$ is the minimal size for an organ to still function and $\tmax$ is the maximal treatment time, for a given treatment, such that the healthy organ is not damaged permanently. \\

Specifically, we need the following clinical information:
\begin{quote}
{\bf Patient/organ specific:}
\begin{enumerate}
\item radiosensitivity parameters of the healthy organ; $\alpha, \beta$.
\item minimal viable size of the organ; $L$
\item initial organ size; $Z(0)$
\item normal organ size; $M$
\item mean organ repair rate if damaged; $\mu$
\end{enumerate}
{\bf Treatment specific:}
\begin{enumerate}
\item treatment schedule, fractionated, hypo-hyper fractionation, brachytherapy etc.; $D(t)$. The radiation sensitivity parameters $\alpha, \beta$ and the treatment schedule $D(t)$ enter the hazard function $h(t)$ (\ref{h:unif}) or (\ref{h:frac}) and hence the removal rate $r(t)$.
\end{enumerate}
\end{quote} 
We solve the logistic equation (\ref{logi2}) until $Z(\tmax)=L$ to find the maximal tolerable treatment time $\tmax$. The corresponding {\it maximal tolerable treatment dose} is  
\[ \Dmax = D(\tmax).\]

As an example we use the parameter values of $\alpha/\beta=3$ from Table \ref{t:para} for three uniform treatments with dose rates $d=2.5, 3, 3.5 $ Gy/day and the two tissue types A and B. The NTCP curves were shown in Figure \ref{compareNTCP} (right). The maximal tolerable dose and the maximal time of exposure in these cases are listed in Table \ref{t:Dmax}.

\begin{table}
\begin{center}
\begin{tabular}{|c|cc|cc|}
\hline
&Tissue A& & Tissue B & \\
dose rate (Gy/day) & $\tmax$ (day) & $\Dmax$ (Gy)& $\tmax$ (day) & $\Dmax$ (Gy) \\
\hline
2.5 & 33 & 82.5 & 15.5 & 38.75 \\
3 & 24 & 72 & 12 & 36\\
3.5 & 17 & 59.5 & 9.5 & 33.25\\
\hline
\end{tabular}
\caption{Maximal treatment time $\tmax$ and maximal tolerable dosages $\Dmax$ for three uniform radiation treatments. tissue A describes a fast repairing tissue, while tissue B is slow repairing.  \label{t:Dmax}}
\end{center}
\end{table} 

\section{Conclusions}\label{s:conclusions}
We introduced a mathematical model for the normal tissue complication probability (NTCP), which is based on patient specific, organ specific and treatment specific parameters. This of course means that we do not provide a one-fits-all formula. Rather, we present a framework such that in a given situation, a NTCP can be derived. The analysis of the stochastic birth-death process suggests to use the logistic differential equation (\ref{logistic}) as a good indicator of the NTCP. Hence the model is no longer   complicated, rather, it is mathematically simple. There is even an explicit solution to the logistic equation. Moreover, the number of parameters that are needed is quite limited ($\alpha, \beta, Z(0), M, L$) and there is real hope that these parameters can be estimated for many healthy tissues in the future.  If this is achieved, we obtain a biologically-based formulation of the NTCP instead of a statistically based NTCP as the one by Lyman, for example. \\

The choice of hazard function $h(t)$ is very important and to illustrate our method we considered two different extreme cases of tissues. For tissue A we assumed that it repairs tissue very quickly, however, single strand breaks can interact disrespective the time they were created, (\ref{h:unif}). Tissue B repairs very slowly and single strand breaks can only interact if they are no more than 6h apart  (\ref{h:frac}). Our method worked very well for these extreme cases, as well as for other combinations which we did not show here. \\

In a next step we can then compare the tumor control probability to the NTCP and formulate a  constraint optimization problem and an optimal control problem. These are based on the following patient specific and organ specific parameters:  
\begin{itemize}  
\item radio sentisitvities of tumor at hand; $\alpha_t, \beta_t$ and initial tumor size $n_0$.
\item radio sensitivities of the involved healthy tissue; $\alpha_h, \beta_h$
\item initial size of the healthy tissue $Z_0$, the normal size of the healthy tissue $M$, and the minimal tolerable size of the organ at hand $L$.
\item tumor growth rate $b$, and tumor death rate $r_t(t)$ based on a radiation schedule $D(t)$
\item healthy tissue repair rate $\mu$, and death rate of healthy tissue due to treatment $r_h(t)$, given by a treatment $D(t)$, which can include a dose-volume histogram of the exposed healthy tissue. 
\end{itemize}

In the following we denote by ${\cal D}$ the set of admissible treatment schedules consisting of certain functions $D:[0,\tmax]\rightarrow \mathbb{R}^+$. Note that ${\cal D}$ can include restrictions on the maximal dose in total and per time interval and further a-priori choices on the type of treatment (e.g. discrete radiation events, continuous radiation, weekends off, etc.). 
For a given treatment schedule $D$ let $(n(\cdot;D), Z(\cdot; D))$ denote the solutions of 
\begin{eqnarray*}
\frac{d n}{dt} &=& (b-r_t(t)) n ,\qquad n(0)=n_0\\
\frac{d Z}{dt} &=& \mu Z \Bigl(1-\frac{Z}{M}\Bigr) - r_h(t) Z, \qquad Z(0) = Z_0.
\end{eqnarray*}
 Note that the variable $D(t)$ enters the radiation induced death rates $r_t(t), r_h(t)$ through the hazard functions.

The goal of the radiation therapy taking into account TCP and NTCP is to achieve tumor control in a certain time interval while restricting the damage. Note that the first means to have
TCP equal to one at final time $\tmax$, which can be expressed already by the formula of Zaider-Minerbo (\ref{ZMTCP}). The constraint has to be formulated in the whole time interval $[0,\tmax]$ however, since too strong damage during the treatment cannot guarantee recovery even if the NTCP is again below a threshold at time $\tmax$.\\

{\bf Control problem:}
Find $D \in {\cal D}$ such that
\[\TCP_{ZM}(\tmax) =  \left(1-\frac{n(\tmax;D)}{n_0 + b n_0 \int_0^\tmax\frac{dt}{n(t;D)} }\right)^{n_0} = 1, \qquad Z(t;D)\geq L \quad\mbox{for all}\quad 0<t\leq \tmax.\]

Since it may be difficult or even impossible  to achieve exact controllability in a finite time interval, we alternatively formulate an optimization problem rather in the tradition of optimizing treatment schedules:\\

{\bf Optimization problem:}
\[ \max_{D \in {\cal D}}\TCP_{ZM}(\tmax)  ,\quad Z(t;D) \geq L \mbox{~for all~} 0<t\leq \tmax.\]

The analysis of this optimization problem and the control problem depends on specific choices of ${\cal D}$ and parameters, which exceeds the scope of this paper and it is an interesting problem for future research.

\bigskip

{\bf Acknowledgements:} Research leading to this paper was carried out when the first author was with WWU M\"unster, partly visiting University of Alberta supported via the PROMOS programm funded by the German Academic Exchange Service (DAAD). Parts of this manuscript were written during a research stay of TH at the Mathematical Biosciences Institute in Columbus, Ohio. TH is supported through NSERC. MB acknowledges support by the German Science Foundation DFG via  EXC 1003 Cells in Motion – Cluster of Excellence, M\"unster, Germany.

\bibliographystyle{plain}
\bibliography{../../../../tex/lit/litdiss}

\end{document}